# Targeting Neurodegeneration: Three Machine Learning Methods for G9a Inhibitors Discovery Using PubChem and Scikit-learn

Mariya L. Ivanova[1,*, ORCID], Nicola Russo[1, ORCID] and Konstantin Nikolic[1, ORCID]
Author affiliations
[1]School of Computing and Engineering, University of West London, London, UK
*Corresponding author mariya.ivanova@uwl.ac.uk

## Abstract

In light of the increasing interest in G9a's role in neuroscience, three machine learning (ML) models, that are time efficient and cost effective, were developed to support researchers in this area. The models are based on data provided by PubChem and performed by algorithms interpreted by the scikit-learn Python-based ML library. The first ML model aimed to predict the efficacy magnitude of active G9a inhibitors. The ML models were trained with 3,112 and tested with 778 samples. The Gradient Boosting Regressor perform the best, achieving 17.81% means relative error (MRE), 21.48% mean absolute error (MAE), 27.39% root mean squared error (RMSE) and 0.02 coefficient of determination (R2) error. The goal of the second ML model called a CID_SID ML model, utilised PubChem identifiers to predict the G9a inhibition probability of a small biomolecule that has been primarily designed for different purposes. The ML models were trained with 58,552 samples and tested with 14,000. The most suitable classifier for this case study was the Extreme Gradient Boosting Classifier, which obtained 78.1% accuracy, 84.3% precision,69.1% recall, 75.9% F1-score and 8.1% Receiver-operating characteristic (ROC). The third ML model based on the Random Forest Classifier algorithm led to the generation of a list of descending-ordered functional groups based on their importance to the G9a inhibition. The model was trained with 19,455 samples and tested with 14,100. The probability of this rank was 70% accuracy.

## 1. Introduction

The epigenome regulator G9a, identified as histone-lysine N-methyltransferase, has been demonstrated to exert a significant influence on the nervous system, modulating vital cellular processes fundamental to neuronal function. For example, a novel mechanism of the Alzheimer's disease (AD) pathogenesis was uncovered, where the histone methyltransferase G9a plays essential role in the regulation the translation of hippocampal proteins, thereby influencing the proteopathic characteristics of AD (Chen et al., 2023). Since in Alzheimer`s disease histone methyltransferases are overexpressed (Park et al., 2022) the role of G9a in the progression of neurodegenerative diseases, particularly AD, is being studied. It was discovered that abnormal histone methylation patterns activity of the G9a leads to dysregulated gene expression that in turn disrupts the transcription of genes crucial for synaptic plasticity, neuronal function, and survival (Bellver-Sanchis et al., 2025). G9a's role in neuronal polarization was identified through several key observations, and a peak in the expression of both, the histone methyltransferase G9a and its major splicing isoform, has been noted at the time of axon formation. So, the inhibition of G9a function would lead to an upregulation of RhoA-ROCK, Rho-associated coiled-coil kinase, activity (Wilson et al., 2020). Promising pharmacological effects of G9a inhibition have been observed not only in AD, but also significantly in Huntington's disease, autism spectrum disorder, Parkinson's disease, and neuropsychiatric disorders (Bellver-Sanchis et al., 2024). The major depressive disorder was associated with the level of G9a in the mood regulating regions of the brain which makes the enzyme a potential target for antidepressant treatment (Park et al., 2022). Moreover, the elevated levels of the G9a enzyme in the dorsal prefrontal cortex were linked to the development of schizophrenia (Chen et al., 2024). The G9a enzyme also catalyses the synthesis of the neurotransmitter acetylcholine limiting the loss of basal forebrain cholinergic neurons caused by ethanol. It was found out that a G9a inhibitor both prevented and reversed the ethanol-induced reduction of choline acetyltransferase (ChAT) (Crews et al., 2023; Hwang & Zukin, 2018).

The chronic neuroinflammation and oxidative stress could yield to neurodegenerative disorders, such as Parkinson's disease, Alzheimer's and Multiple Sclerosis (Adamu et al., 2024). On the other hand, neuroinflammation was revealed to strongly induce the G9a-catalyzed repressive epigenetic mark (Rothammer et al., 2022). This, in turn, suppresses the expression of anti-ferroptosis genes, ultimately triggering ferroptosis, an iron-dependent form of programmed cell death. Both autoimmune encephalomyelitis and human multiple sclerosis were examined and identified G9a as a potential therapeutic target for inflammation-induced neurodegeneration. A previously unrecognized function of G9a in protein-specific translation was identified that could be used to treat chronic inflammatory diseases associated with endotoxin tolerance (Muneer et al., 2023).

The G9a enzyme is not associated only with the neurodegeneration diseases. The nerve injury causes dysfunction of potassium (K+) channels in the dorsal root ganglion (DRG), and the exploration of the epigenetic mechanism behind this process revealed that the G9a enzyme plays a key role in suppressing the activity of K+ channel genes and contributes to the transition from acute to chronic pain after nerve injury (Laumet et al., 2015). It has been observed that signs of neuropathic pain in animal models have been abolished when modulators of voltage-gated channels (Na+, Ca2+, K+) were manipulated. However, the expected clinical endpoint has not been achieved. Nevertheless, the voltage-gated K+ channels remain a subject of investigation because blocking or genetic deletion of G9a, which controls the expression of several voltage-gated K+ channels reduces neuropathic hyperalgesia, a state of increased sensitivity to pain (Ghosh and Pan, 2022). Furthermore, G9a was pointed as a potential target for alleviating persistent inflammatory pain because while G9a normally suppresses the expression of Transient Receptor Potential Ankyrin 1 (TRPA1) **a**nd Transient Receptor Potential Vanilloid 1 (TRPV1) ion channels, it surprisingly enhances their expression during tissue inflammation which in turn cause upregulation in the DRG (Ghosh et al., 2025). It diminishes the gene expression of the opioid and cannabinoid nociceptive receptors as well (Ghosh and Pan, 2022; Luo Y et al, 2020). Last, but not least, the transcriptional repressor of neuronal genes outside the nervous system, i.e. the Neuron-Restrictive Silencer Factor (NRSF) protein, recruits G9a in silencing of neuronal genes expression in nonneuronal cells (Roopra et al, 2004).

Different ML models have been developed to serve researchers in Neuroscience, such as ML model based on transcriptomic (the study of all RNA molecules in a cell or tissue) data developed to identify new genes associated with vascular dysfunction in the middle temporal gyrus of brains affected by Alzheimer's disease (Wang et al., 2025); implementation of neuroimaging and ML for achieving of multiclass diagnosis of neurodegenerative diseases (Singh et al, 2019); ML classification based on the carbon 12-isotop 13C Nuclear Magnetic Resonance (NMR) spectroscopy data derived by the Simplified Molecular Input Line Entry System **(**SMILES) notations (Ivanova, Russo and Nikolic, 2025b)

The present study developed three ML models, offering time and cost efficiency approaches, intending to support drug discovery research in neuroscience. The bioassays, PubChem AID 504332 focused on G9a inhibition (PubChem, 2011) and PubChem AID 1996 on the water solubility of small molecules (PubChem, 2010) and all additional data utilised by the three ML models were provided by PubChem, the biggest freely available chemical data worldwide (PubChem, 2025a). In the study were used algorithms interpreted by scikit-learn library for ML (Pedregosa et al., 2011). The ML models were developed in the Jupiter notebook environment (Jupiter, 2024).

The first ML model aimed to predict the magnitude of the active G9a inhibitors. For this purpose, the dataset generation methodology from a study (Ivanova, Russo, Djaid and Nikolich, 2024) specifically focused on ML applications for G9a inhibitors was utilized. Additionally, nine new ML attributes were engineered based on the considered small biomolecules' atomic properties. Given the nature of the considered ML model, only the small biomolecules exhibited G9a inhibition were included in the datasets. Also, since the ML model is a regression, skewness was taken into consideration due to its potential influence on the model's validity. The methodology is presented in Figure 1.

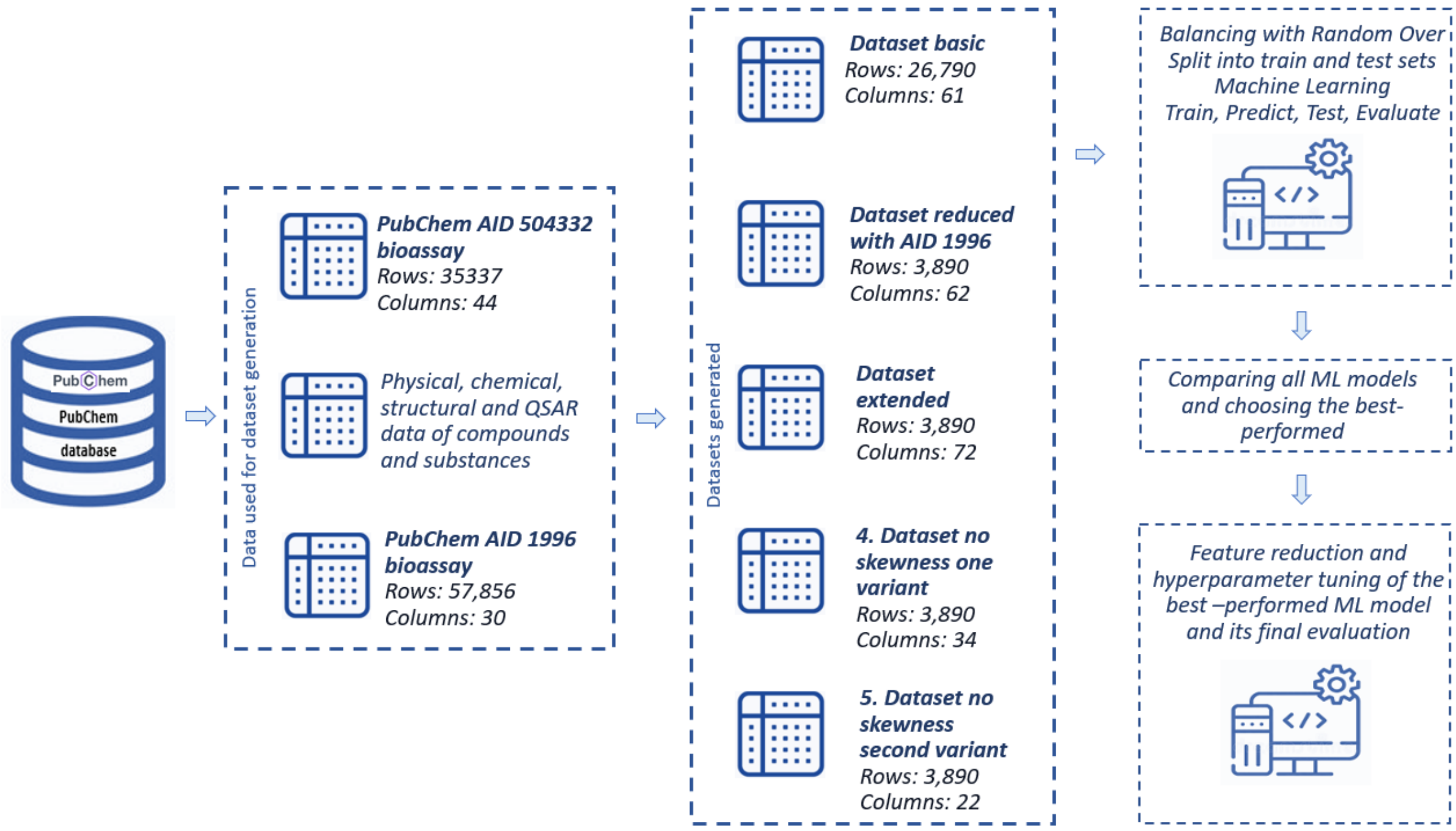

Figure 1.
Methodology of developing an ML model that can predict the magnitude of the G9a inhibitor`s efficacy

Using established methodologies originally tested with different case studies, the second- and third-ML models were developed to explore the potential of these approaches in the context of G9a inhibition. The second model, based on the CID_SID ML model methodology (Ivanova, Russo and Nikolic, 2025a), would allow researchers to check if their compound is a G9a inhibitor, using only its PubChem (PubChem, 2025b) identifiers, i.e. PubChem CID and SID. Thus, a compound initially designed for another purpose could be easily assessed whether it also possesses the potential of being a G9a inhibitor, which, in turn, could accelerate the early stage of drug discovery. Although the identifiers are generally not used for training and testing of ML models, PubChem identifiers make an exception because they have been generated based on chemical structure and similarities among compounds and substances (Kim et al., 2016). The third ML model implemented the information encoded in the small biomolecules generated by the International Union of Pure and Applied Chemistry (IUPAC) nomenclature (IUPAC, 2025). Although the ML model predicted whether a small biomolecule is a G9a inhibitor based on the IUPAC information, this approach primarily aimed to produce a ranked list of functional groups according to their importance and thus speeding up the initial drug discovery process (Ivanova, Russo and Nikolic, 2025c).

## 2. Methodology

### 2.1. Regression ML model for predicting the magnitude of G9a inhibitor efficacy

Initially a dataset was created by extracting Efficacy, CID, SID, Outcomes and Phenotype columns from PubChem AID 504332 (PubChem, 2011). The Efficacy column was then designated as the target variable, and the remaining extracted columns were used to construct the dataset required for this study. The identification of active G9a inhibitor samples was achieved by keeping only the row containing values, as follows: "Active" in the Outcomes column and "Inhibitor" in the Phenotype column. The CIDs and SIDs of the remained samples then were used for downloading data from PubChem (PubChem, 2025b) and built dataset *basic*. Detailed information regarding feature types and derived compound properties, including the computational methodologies employed, can be found in the parent study (Ivanova, Russo, Djaid and Nikolic, 2024) or in the code provided on GitHub.

Dataset basic then was merged with the dataset of the bioassay PubChem AID 1996 (PubChem, 2010) on small molecule keeping only the common for both datasets samples. The resulting dataset was named *reduced.*

For the next dataset, nine new features were generated and added as columns to the reduced dataset. The new dataset was called *extended*. Three of these new features compared the similarity between the SMILES notations of the compounds of the *reduced* dataset and proven G9a inhibitors provided by PubChem:

(i) 6-methoxy-N-methyl-2-(5-methylfuran-2-yl)-N-(1-methylpiperidin-4-yl)-7-(3-pyrrolidin-1-ylpropoxy) quinolin-4-amine (PubChem, 2024)
(ii) Bix-01294 (PubChem, 2009)
(iii) N-[2-[[4-[(1-benzylpiperidin-4-yl) amino]-7-methoxyquinazolin-2-yl] amino]ethyl]-N'-hydroxyhexanediamide;hydrochloride (PubChem, 2025c)

The rest of the new features were engineered with already calculated by the parent study (Ivanova, Russom Djaid and Nikolic, 2024) features. The features obtained were:

(i) Mean value of the relative ratio of Carbon atoms raised to a power equal to the length of the carbon chain of the compound along the 3D axis.

(ii) Mean value of the relative ratio of Carbon atoms raised to a power equal to the relative ratio of the Carbon chain sizes along the 3D axes.

(iii) Mean value of the mass proportion of the carbon atoms raised to a power equal to Carbon atoms chain skewness.

(iv) Mean value of the relative ratio of Hydrogens atoms raised to a power equal to the length of the molecules of the compound along the 3D axes.

(v) Mean value of the relative ratio of Hydrogen atoms raised to a power equal to the relative ratio of the molecule size along the 3D axes.

(vi) Mean value of the mass proportion of the Hydrogen atoms raised to a power equal to the molecule skewness along the 3D axes.

For more details about the new features and their calculation, please refer to the code provided on the GitHub.

To address the impact of skewness on the regression model, two datasets were created without it. The first dataset contained features with skewness of ± 0.5% and was called a dataset *without skewness*. The second dataset was generated also from columns which contained data with skewness smaller than ± 0.5 %, but after the entire data was log transformed. It was called a dataset *without skewness after log transformation*.

Each dataset was complete and numerical. After splitting into features and targets, features were standardized using Standard Scaler to prevent feature prioritization. Then entire data was split in sets for training and testing. The training parts were balanced in order to avoided bias towards the majority class. The prepared test and train set then were used for ML with the next estimators:

(i) Decision Tree Regressor: the algorithm starts by selecting the best feature to split the data based on a certain criterion; splitting based on the chosen features and a specific split point, creating two branches from the node; recursion, repeated for each subset and continue until a stopping criterion is met; at a leaf node, the model predicts a continuous value, typically the average of the target values observed in the training data associated with that node.

(ii) Random Forest Regressor: Instead of relying on a single decision tree, a Random Forest constructs multiple decision trees, each trained on distinct subsets of both the data and features. The aggregate prediction is determined by averaging the predictions of all individual trees. It creates numerous decision trees using varied data and feature subsets. The final prediction is the average of these trees' output.

(iii) Gradient Boosting Regressor: Unlike Random Forests, which build trees independently, Gradient Boosting builds trees sequentially, each correcting the errors of the previous one.

(iv) Support vector Regressor: Unlike traditional linear regression, SVR can model non-linear relationships by mapping the data into a higher-dimensional space using kernel functions. The goal is to find a hyperplane in this higher-dimensional space that best fits the data, with the idea of maximizing the margin while minimizing the error.

Each one of these regressors was used for training, predicting and evaluating of the ML model according to the best practice recommended in the ML literature (Wang et al. 2020). Feature reduction was applied on the best performed estimator. The ML model metrics, such as MAE, RMSE and R2 were compared and conclusions drown.

## 2.2. CID_SID ML model to classify a small biomolecule based on its PubChem identifiers.

For this model CIDs, SIDs, outputs and phenotypes of the small biomolecules considered in the PubChem AID 504332 (PubChem, 2011) bioassay were extracted. Similarly to the data generation approach mentioned above, the values in columns with the outputs and phenotypes were used. However, this time, the samples possessing the values "Active" and "Inhibitor" in the specified columns were labelled 1, and those with "Inactive" and "Inhibitor" were assigned 0. Thus, a dataset was obtained comprising only these two labels, corresponding to the activity of the G9a inhibitors, and their relevant CIDs and SIDs.

Since the ML model was classification, the severe misbalancing of the dataset had to be handled. For this purpose, the amount of the inactive compounds was decrease through two stages. Initially, the G9a inhibitor dataset was merge with the PubChem AID1996 bioassay dataset (PubChem, 2010). The resulting reduced inactive samples were concatenated to the full set of the active inhibitors compounds. The dataset then was split to testing and training sets. The testing sets were ensured to have equal number of each class. The remaining samples, created the training set, were finally balanced with random over sampler (Imbalanced-learn, 2024) which randomly selected samples from the majority class, filling the minority class until both classes were with equal number of samples. The resulting datasets, the testing and training sets, were used for developing ML models with Decision Tree Classifier (DTC), Random Forest Classifier (RFC), Gradient Boosting classifier (GBC), Extreme Gradient Boosting classifier (XGBC), Support Vector Classifier (SVC) algorithms. The difference between these classifiers and their corresponding regressors listed above consists in the type of target variable they handle. The target variable for the classifiers is a categorical data type, whereas for the regressors is continuous. The CID_SD ML models were evaluated with the relevant metrics, appropriate for the classification ML models such as Accuracy (the percentage of the correct predictions), Precision (the percentage of how many of the positive predicted samples were actually positive), Recall (Percentage of how many of the positive samples in the dataset were predicted accurately), F1 (The harmonic mean between precision and sensitivity ) and ROC (the proportion between the true and false positive predictions, when the threshold varied). The best performed algorithm then was hyperparameter tuned by Optuna (Akiba, 2019). To visualize the final CID_SID machine learning model, the confusion matrix, learning curve, AUC were plotted and the classification report displayed. For more details, please refer to the parent CID_SID ML model study (Ivanova, Russo, and Nikolic, 2025a) and/or the code repository of the current study on GitHub.

### 2.3. IUPAC based ML model for providing a descendent ordered list of functional groups according to their importance to G9a inhibition

In this case, the IUPAC names of the small biomolecules considered in the PubChem AID 504332 (PubChem, 2011) bioassay were parsed into strings containing four or more letters. Each one of these strings was used for generation of a column in a data frame. In these columns the presence of a relevant functional group was noted with 1 and the absence with 0. These data were ladled and used for ML. The reduction of inactive compounds, balancing of data, ML training, predicting and evaluating resemble the approach of CID_SID ML model presented about. For more details, please, refer to the IUPAC based source article (Ivanova, Russo, and Nikolic, 2025c) and/or the code repository of the current study on GitHub. Given the large number of features, Principal Component Analysis (PCA) was performed to reduce dimensionality, with the expectation of improving the machine learning model's performance (Costa et al., 2024).

## 3. Results and discussion

### 3.1. A regression ML model predicting the magnitude of G9a inhibitor`s efficacy

As a result of applying the methodology described above, the first dataset, called *basic*, contained 26,790 rows and 61 columns; the second dataset, called *reduced*, contained 3,890 rows and 62 columns; the third dataset, called *extended*, contained 3,890 rows and 72 columns; the fourth dataset, called *without skewness*, contained 3,890 and 34 columns; the fifth dataset, called *without skewness after log transformation*, contained 3,890 and 22 columns (Figure 1). For more details, please see the code provided on GitHub and the feature generation parent study (Ivanova, Russo, Djaid and Nikolic, 2024)

The ML results for the estimators and datasets specified above are available on the Electronic Supplementary Material, as follows: Table ESM1, Table ESM3, Table ESM5, Table ESM7, Table ESM9 are with metrics of the ML models used the datasets *basic, reduced, extended, without skewness, without skewness after log transformation*, and Table ESM2, Table ESM4, Table ESM6, Table ESM8 present the five-fold cross-validation results for the same machine learning models, applied to the datasets in the same order. The evaluation of the ML models was visualised by Figure 2 that compares the MAE across different datasets, and Figure 3 which displays the corresponding relative errors. Amongst the considered ML models and datasets, the GBR estimators with the dataset *without skewness* performed the best, achieving 20.17% MAE on train sets, 21.29% MAE on test sets; 25.51% RMSE on train sets, 27. 03% RMSE on test sets; 0.25 R2 with train sets, and 0.08 R2 with test sets, followed by the RFR with 8.55% MAE on train sets, 21. 42% MAE on test sets; 10.95% RMSE on train sets, 27. 38% RMSE on test sets; 25.51% RMSE on train sets, 27. 03% RMSE on test sets; 0.25 R2 with train sets, and 0.08 R2 with test sets (Table ESM7 ). Amongst the results, the big difference between the RFR errors accomplished by the train and test sets was an indicator that the ML model is overfitted. Moreover, the small R2 metric of GBR, i.e. 0.002, was an indicator that this estimator behaved more like a statistical model than ML one, enabled to learn relationships between independent and dependent variables. The five-fold cross-validation, scoring with MAE, which estimated how well GBR would perform on unseen data, obtained cross-validation score 23.06 with 0.85 standard deviation, which is relatively close to the training sore, so GBR was picked up for further development.

Applying the mentioned above two feature reduction algorithms generated two lists of descending order of the features, scored according to their influence (Figure ESM1, Figure ESM2) However, exploring the influence on each feature showed that the influence of the number of the features is not significant. It turned out that the first three features in both lists were identical: Topological Polar Surface Area (TPSA), referring to the sum of polar atom/molecule surface contributions (Ertl, Rohde, & Selzer, 2000), water solubility at pH 7.4 (PubChem, 2010) and XLogP3, the octanol-water partition coefficient (Cheng, et al., 2007). These features were used for generation of a new dataset, which was later utilized by GBR and obtained 21.37% MAE on train sets, 21.48% MAE on test sets; 27.08% RMSE on train sets, 2.39% RMSE on test sets; 0.16 R2 with train sets, and 0.06 R2 with test sets. Furthermore, the hyperparameter tuning recommended max_depth=3, n_estimators=200, min_samples_split=2, min_samples_leaf=1, learning_rate=0.01 like the best option. However, performing the GBR with the hyper parameter tuned values slightly increased the errors to 22.49% MAE on train sets, 21.64% MAE on test sets; 28.477% RMSE on train sets, 27.56% RMSE on test sets; 0.07R2 with train sets, and 0.04 R2 with test sets. While hyperparameter tuning was conducted using the Grid Search algorithm, the implementation of other techniques would likely improve the machine learning mode. However, despite the potential benefits of extensive hyperparameter tuning, the observations made during the study led to the conclusion that further research into feature engineering is necessary to minimize errors.

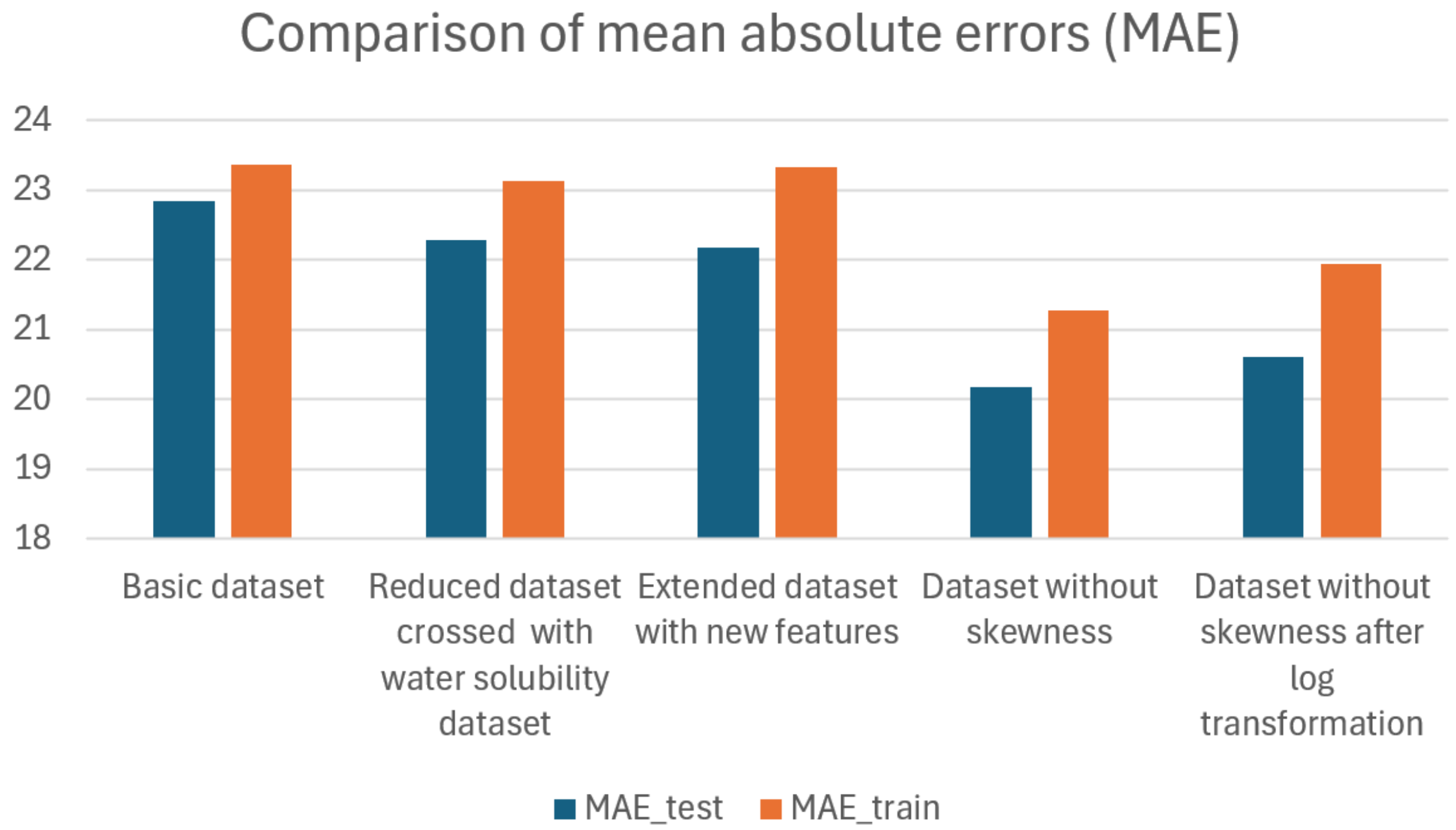

Figure 2. Comparison of mean absolute errors among the five datasets obtained by the best performed estimators: the first, fourth and fifth results are achieved by Gradient Boosting Regressor; the second and third results are achieved by Support Vector Regressor.

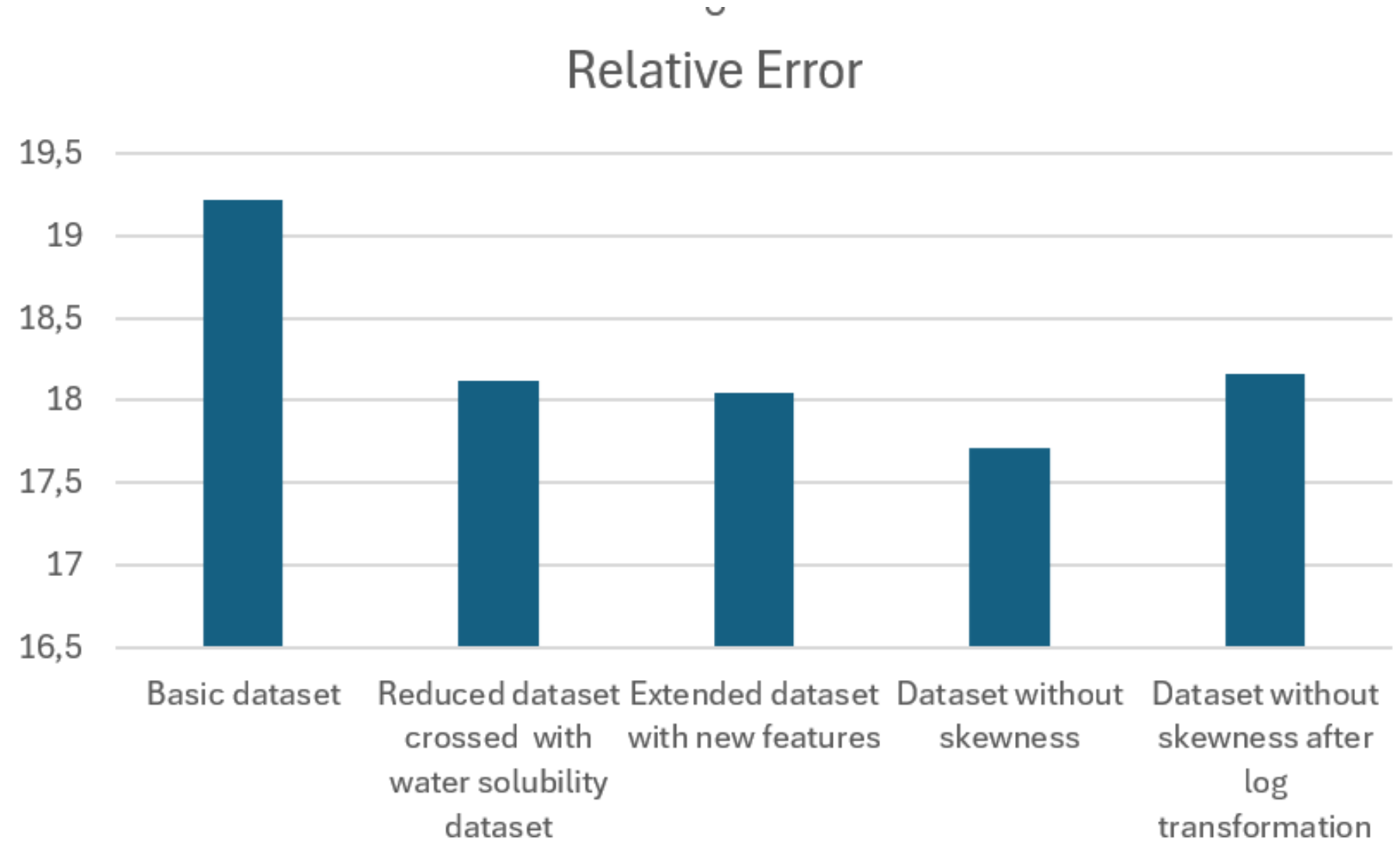

Figure 3. Relative error of ML models performed with the five datasets.

## 3.2. CID_SID ML model predicting whether a small biomolecule is a G8a inhibition based on the PubChem identifiers

As a result of the dataset generation methodology explained above, a dataset containing 36276 inactive compounds, ladled as 0, and 30,242 active compounds labelled as 1 was obtained. From these samples, 7,000 random compounds from each class were used for testing and the rest were balanced and the resulting 58,552 samples were used for training, From the listed classifier above, XGBC performed the best, achieving 78.9% Accuracy, 85.8% Precision, 69.4% Recall, 76.7% F1-score and 79% ROC, followed by RFC with 78.4% Accuracy, 82.7% Precision, 71.7% Recall, 76.8% F1-score and 78.4% ROC (Table 1)

Table1. CID_SID ML model

| Algorithm | Accuracy | Precision | Recall | F1 | ROC |
|---|---|---|---|---|---|
| XGBoost | 0.789 | 0.858 | 0.694 | 0.767 | 0.790 |
| RandomForest | 0.784 | 0.827 | 0.717 | 0.768 | 0.784 |
| GradientBoost | 0.777 | 0.894 | 0.629 | 0.738 | 0.777 |
| K-nearest | 0.762 | 0.779 | 0.731 | 0.754 | 0.762 |
| Decision | 0.750 | 0.763 | 0.724 | 0.743 | 0.750 |
| SVM | 0.729 | 0.871 | 0.538 | 0.665 | 0.729 |

The cross-validation score for XGBC with five-fold cross-validation was 0.7998 with 0.0013 standard deviation and 0.7912 cross validation score with 0.0036 standard deviation for RFC (Table ESM 11). The scrutinised for overfitting of the CID_SID ML model revealed that the overfitting, i.e. the deviation between the test and train accuracy is higher than 5 %, started after max_depth=9, where the test accuracy was 78.7%, and the train accuracy was 83.2% (Figure ESM3). The hyperparameter tuning with Optuna with 100 trials (for details about the range of hyperparameters, please, see the code provided on GitHub) didn't improve the model. The hyperparameter tuned XGBC tuned by Optuna achieved 77.9% Accuracy, 84.7% Precision, 68.2% Recall, 75.5% F1-score and 75.5% ROC. So, the final ML model remained with the default for the XGBC hyperparameters values and achieved 78.7% Accuracy, 84% Precision, 70.8% Recall, 76.8% F1-score and 78.7% ROC with max_depth=9.

The classification report of the final CID_SID ML model predi6cting whether a compound is a G9 inhibitor based solely on its CID and SID in provided in Table 2. The CID_SID ML model is visualised by its learning curve (Figure ESM4), the area under the curve (AUC) (Figure ESM5) and the matrix (Figure ESM6).

Table 2. The CID_SID Random Forest Classification ML model classification report

| | precision | recall | f1-score | support |
|---|---|---|---|---|
| Active (target 1) | 0.75 | 0.87 | 0.80 | 7000 |
| Inactive (target 0) | 0.84 | 0.71 | 0.77 | 7000 |
| accuracy | | | 0.79 | 14000 |
| macro avg | 0.79 | 0.79 | 0.79 | 14000 |
| weighted avg | 0.79 | 0.79 | 0.79 | 14000 |

### 3.3. IUPAC based ML approach for providing a descendent order list of functional groups according to their importance regarding G9a inhibition model predicting whether a small biomolecule is a G8a inhibition bas

Parsing the IUPAC notations of the compounds considered in the PubChem AID 504332 (PubChem, 2011) bioassay was a prerequisite for generation of a data frame with 63,792 rows with samples and 4,927 columns with features, which corresponded to the parsed IUPAC names. In this dataset, 33,555 samples were labelled as inactive and 30,237 as an active G9a inhibitors. As a result of data preprocessing, test-train splitting and balancing of the training set, the ML model was trained with 54,310 samples and tested with 12,800 samples. RFC, used in this case study, achieved 77.1% Accuracy, 82.4% Precision, 68.9% Recall, 75.1% F1-score and 77.1% ROC and five-fold cross-validation score of 0.7953 with 0.0032 standard deviation. The scrutinising for overfitting revealed that the overfitting occurred at max_depth=25, where the train accuracy was 75.8% and the test accuracy was 70.6%, i.e. the deviation between them was bigger than 5%. The hyperparameter tuning with 10 trials did not improve the model, so similarly to the CID_SID ML model, the IUPAC based model kept the default hyperparameter for RFC and achieved 70.5% Accuracy, 78.1% Precision, 56.9% Recall, 65.8% F1-score and 70.5% ROC with max_depth=24.The five-fold cross-validation score of this ML model was 0.7104 with standard deviation of 0.0049 implied that the model was robust and reliable.

The classification report of the ML model predicting whether a compound is a G9 inhibitor based on the information encoded in the IUPAC names in provided in Table 3. The ML model results are visualised by plotting its matrix (Figure ESM8).

Table 3. The IUPAC based Random Forest classification ML model

| | precision | recall | f1-score | support |
|---|---|---|---|---|
| Active (target 1) | 0.66 | 0.84 | 0.74 | 6400 |
| Inactive (target 0) | 0.77 | 0.56 | 0.65 | 6400 |
| accuracy | | | 0.70 | 12800 |
| macro avg | 0.72 | 0.70 | 0.69 | 12800 |
| weighted avg | 0.72 | 0.70 | 0.69 | 12800 |

The descendent orders of the functional groups obtained by RFC feature importance (Scikit-learn, 2025b) and SelectKBest with chi-2 statistics approach (Scikit-learn, 2025c) algorithms are illustrated in Figure 4 and Figure 5. Comparing the two lists revealed that although the order of both is slightly different the functional groups are nearly identical for both orders. The only differences were for the first list contained methyl, amino and ethyl, unlike the second list which contained pyrrrol, diazinane and methylfuran. Despite the computational nature of the approach, it was hypothesized that such reductions of the functional group would provide practical insights into drug discovery related to G9a inhibition.

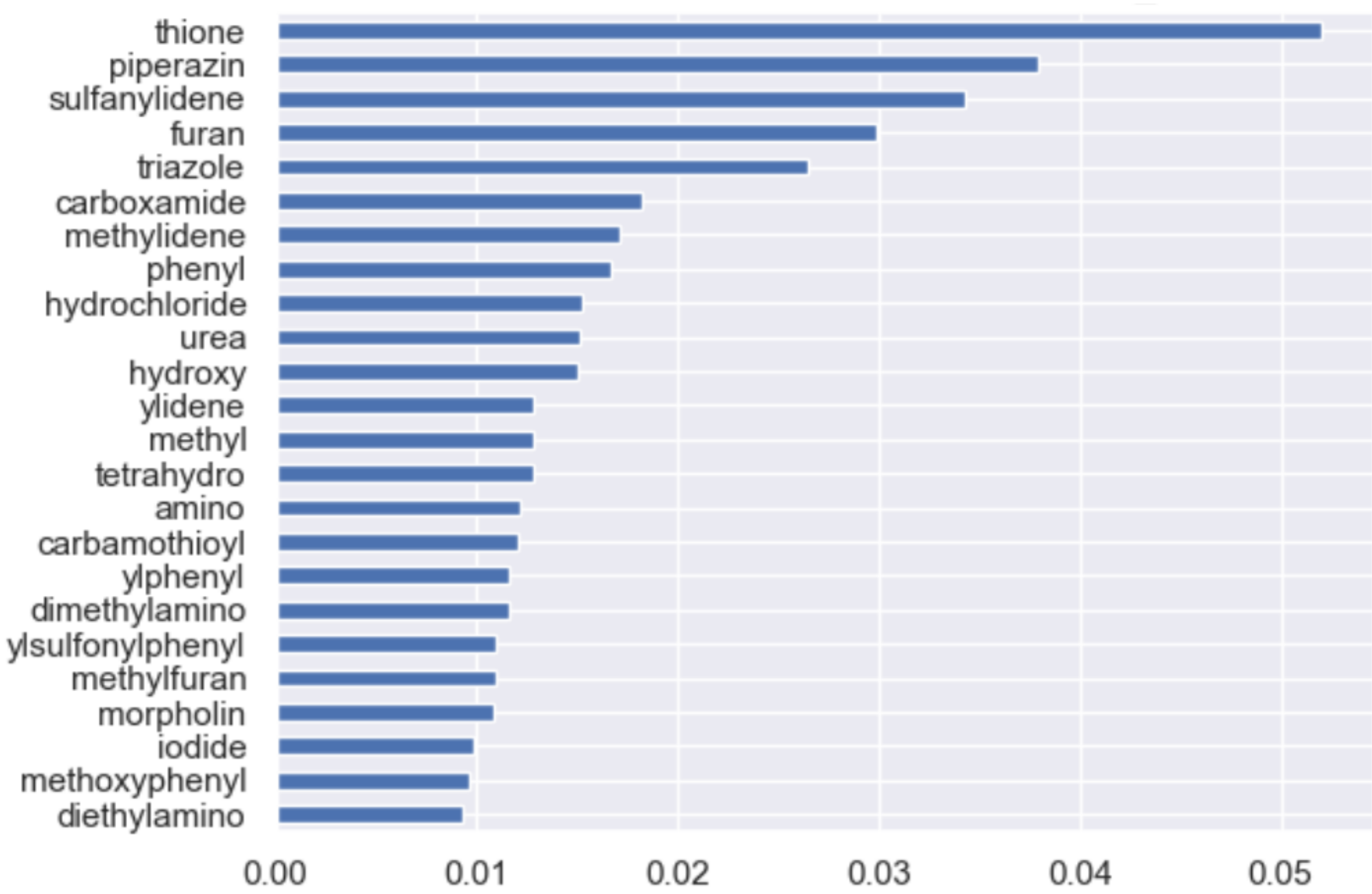

Figure 4. Feature importance for RFC ML model
focused on the G9a inhibition

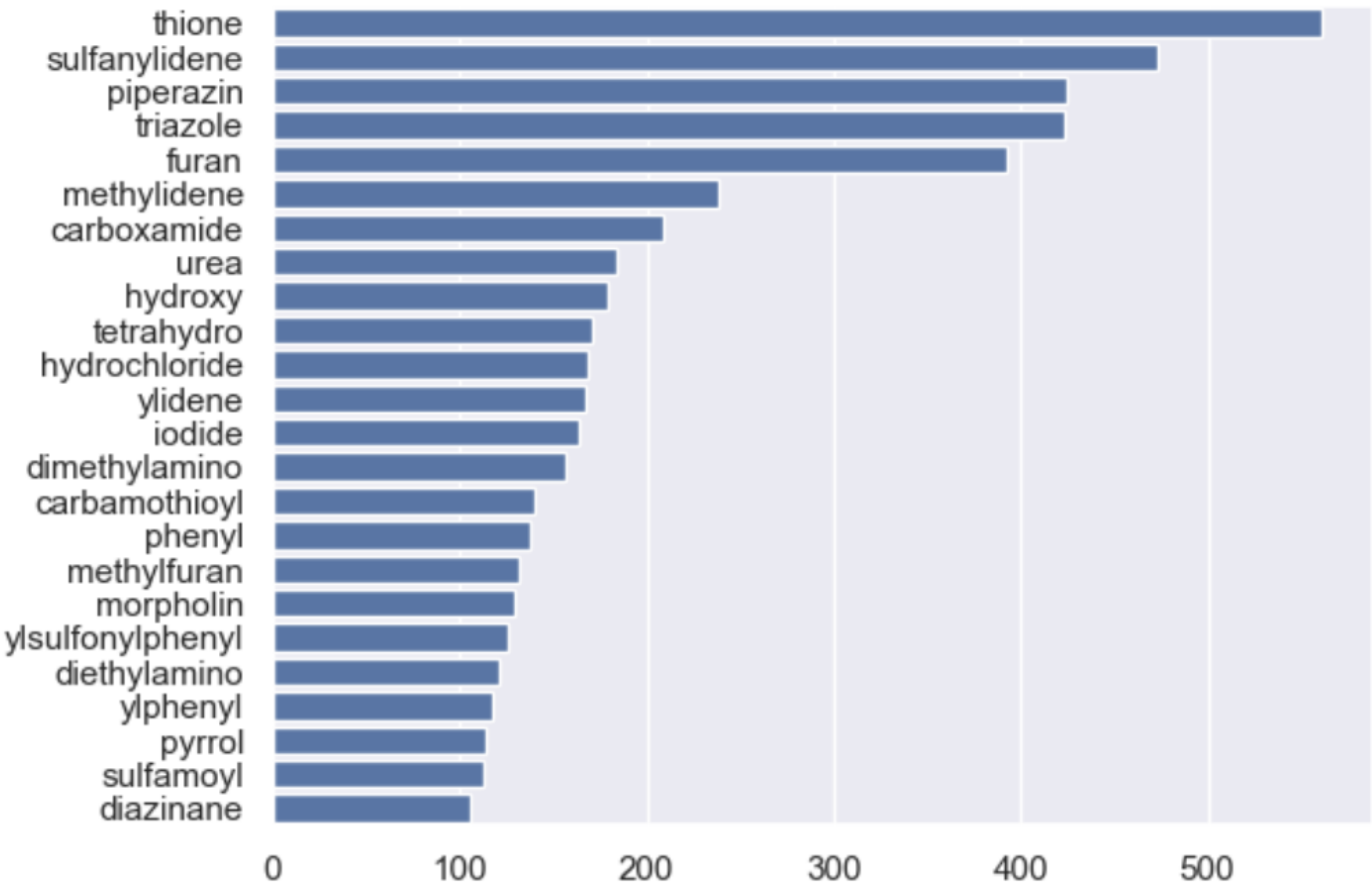

Figure 5. Feature importance based on the Scikit-Learn SelectKBest function with chi2 statistical approach, focused on the G9a inhibition

## Conclusion

The first ML model developed in the study predicts the efficacy magnitude of G9a inhibitors. It is time and cost-efficient compared to the time and cost needed for laboratory experiments. The ML model is based on three features, TPSA, XLogP3 and water solubility at pH 7.4, which were chosen mainly because these features are relatively easy to obtain compared to the rest. However, the observations revealed that the combinations of other features considered in the study also could be similar to the results achieved by the final ML model. So, the researchers by themselves have the option to choose which feature, from the given ones, would be most suitable for them and implemented them subsequently for the ML prediction. The study needs to explore other anatomic-based features that would reduce the errors. The second and the third ML models provide approaches which are expected to reduce the time and cost necessary for the initial stage of drug discovery with respect to G9a inhibition. About the generation of the descending orders based on the ML model with the IUPAC data, expectedly, different machine learning runs produced different results. However, the differences between the results were not significant. So, the interested researchers could use the approach to achieve statistics for more precise identification of the importance of the functional groups with respect to G9a inhibition. The ML models were hyperparameter-tuned by Optuna and did not achieve better results. However, increasing the number of trials and expand the scope of the hyperparameter values implies that the ML model would improve their performance.

## Author Contributions

MLI, NR and KN conceptualized the project and designed the methodology. MLI and NR wrote the code and processed the data. KN supervised the project. All authors were involved with the writing of the paper.

## Acknowledge

MLI thanks the UWL Vice-Chancellor's Scholarship Scheme for their generous support. We sincerely thank NIH for providing access to their PubChem database. The article is dedicated to Luben Ivanov.

This preprint has not undergone peer review or any post-submission improvements or corrections. The Version of Record of this article is published in the Journal of Computer-Aided Molecular Design and is available online at https://doi.org/10.1007/s10822-025-00642-z

## Data and Code Availability Statement

The raw data used in the study is available through the PubChem portal: https://pubchem.ncbi.nlm.nih.gov/

The code generated during the research is available on GitHub: https://github.com/articlesmli/IUPAC_ML_model_TDP1.git

## Conflicts of Interest

The authors declare no conflict of interest.

References

Adamu, A., Li, S., Gao, F. & Xue, G. (2024). The role of neuroinflammation in neurodegenerative diseases: current understanding and future therapeutic targets. *Frontier in Aging Neuroscience*, *16*(4), 1347987. https://doi.org/10.3389/fnagi.2024.1347987

Akbar, S., Ullah, M. Raza, A.  Zou, Q.  and Alghamdi, W. (2024) DeepAIPs-Pred: Predicting Anti-Inflammatory Peptides Using Local Evolutionary Transformation Images and Structural Embedding-Based Optimal Descriptors with Self-Normalized BiTCNs, *J. Chem. Inf. Model., 64* (24), 9609-9625. https://doi.org/10.1021/acs.jcim.4c01758

Akiba, T., Sano, S., Yanase, T., Ohta, T. & Koyama, M. (2019) Optuna: A Next-generation Hyperparameter Optimization Framework. ArXiv. https://doi.org/10.48550/arXiv.1907.10902

Alles, S.R.A., & Smith, P.A. (2021). Peripheral Voltage-Gated Cation Channels in Neuropathic Pain and Their Potential as Therapeutic Targets. *Frontiers in Pain Research*, *2*(12). https://doi.org/10.3389/fpain.2021.750583

Bellver-Sanchis, A., Ávila-López, P., Tic, I., Valle-García, D., Ribalta-Vilella, M., Griñán-Ferré, C., et al. (2024). Neuroprotective effects of G9a inhibition through modulation of peroxisome-proliferator activator receptor gamma-dependent pathways by miR-128. *Neural Regeneration Research*, *19*(11), 2532-2542. https://doi.org/10.4103/1673-5374.393102

Bellver-Sanchis, A., Ribalta-Vilella, M., Irisarri, A., Gehlot, P., Choudhary, B.S., Griñán-Ferré, C., et al. (2025), G9a an epigenetic therapeutic strategy for neurodegenerative conditions: from target discovery to clinical trials. *Medicinal Research Reviews*.  https://doi.org/10.1002/med.22096

Chen, X., Xie, L., Sheehy, R., Xiong, Y., Muneer, Deshmukh, M., et al. (2023) Novel brain-penetrant inhibitor of G9a methylase blocks Alzheimer's disease proteopathology for precision medication. *Research Square*.  https://doi.org/10.21203/rs.3.rs-2743792/v1

Chen, Y-Z., Zhu X-M., Lv, P., Hou, X-K. Ying. P., Yao, J et al. (2024). Association of histone modification with the development of schizophrenia. *Biomedicine & Pharmacotherapy, 175*(6), 116747, https://doi.org/10.1016/j.biopha.2024.116747

Cheng, T., Zhao, Y., Li, X., Lin, F., Lai L., et al. (2007). Computation of octanol-water partition coefficients by guiding an additive model with knowledge. *Journal of Chemical Information and Modelling, 47*(6), 2140-2148. https://doi.org/10.1021/ci700257y

Costa, A. P. de A., Choren, R., Pereira, D. A. de M., Terra, A. V., Costa, I. P. de A., Moreira, M. Â. L. et al. (2024). Integrating multicriteria decision making and principal component analysis: a systematic literature review. *Cogent Engineering, 11*(1). https://doi.org/10.1080/23311916.2024.2374944

Crews, F.T., Fisher, R.P., Qin, L, & Vetreno R.P. (2023). HMGB1 neuroimmune signalling and REST-G9a gene repression contribute to ethanol-induced reversible suppression of the cholinergic neuron phenotype. Molecular Psychiatry, 28(12), 5159-5172. https://doi.org/10.1038/s41380-023-02160-6

Ertl, P., Rohde, B., & Selzer, P. (2000). Fast calculation of molecular polar surface area as a sum of fragment-based contributions and its application to the prediction of drug transport properties. *Journal of Medicinal Chemistry, 43*(20), 3714-7. https://doi.org/10.1021/jm000942e

Ghosh, K., & Pan, H.L. (2022). Epigenetic mechanisms of neural plasticity in chronic neuropathic pain. *ACS Chemical Neuroscience,* 13 (4), 432-441. https://doi.org/10.1021/acschemneuro.1c00841

Ghosh, K., Huang, Y., Jin, D., Chen, S.R., & Pan, H.L. (2025). Histone methyltransferase G9a in primary sensory neurons promotes inflammatory pain and transcription of Trpa1 and Trpv1 via bivalent histone modifications. *Journal of Neuroscience, 45* (6), e1790242024. https://doi.org/10.1523/jneurosci.1790-24.2024

Hwang, J.Y., & Zukin, R.S. (2018) REST, a master transcriptional regulator in neurodegenerative disease. *Current Opinion in Neurobiology, 48*, 193-200. https://doi.org/10.1016/j.conb.2017.12.008

Imbalanced Learn (2025) *RandomOverSampler.* https://imbalanced-learn.org/stable/references/generated/imblearn.over_sampling.RandomOverSampler.html Accessed 20 February 2025

IUPAC (2025). Home page. *IUPAC.* https://iupac.org/ Accessed 20 February 2025

Ivanova, M. L., Russo, N., Djaid, N., & Nikolic, K. (2024). Application of machine learning for predicting G9a inhibitors. Digital Discovery 3(10): 2010-2018. https://doi.org/10.1039/D4DD00101J

Ivanova, M. L., Russo, N., & Nikolic, K. (2025a). Predicting novel pharmacological activities of compounds using PubChem IDs and machine learning (CID-SID ML model). *ArXiv.* https://doi.org/10.48550/arXiv.2501.02154

Ivanova, M. L., Russo, N., & Nikolic, K. (2025b). Leveraging 13C NMR spectrum data derived from SMILES for machine learning-based prediction of a small molecule functionality: a case study on human Dopamine D1 receptor antagonists. *ArXiv.* https://doi.org/10.48550/arXiv.2501.14044

Ivanova, M. L., Russo, N., & Nikolic, K. (2025c). Hierarchical Functional Group Ranking via IUPAC Name Analysis for Drug Discovery: A Case Study on TDP1 Inhibitors *ArXiv* https://doi.org/10.48550/arXiv.2503.05591

Jupyter (2024). *Home page*. *Jupyter.* https://jupyter.org/ Accessed 4 Jan 2025

Kim, S., Thiessen, P.A., Bolton, E.E., Chen, J., Fu, G., Gindulyte, A., Han, L., He, J., He, S., et al. (2016). PubChem Substance and Compound databases. *Nucleic Acids Research,* 44, D1202-13. https://doi.org/10.1093/nar/gkv951

Laumet, G., Garriga. J, Chen, S.R., Zhang, Y., Li, D.P, Pan, H.L., et al. (2015). G9a is essential for epigenetic silencing of K (+) channel genes in acute-to-chronic pain transition. *Nature Neuroscience*,*18*(12), 1746-55. https://doi.org/10.1038/nn.4165

Luo, Y., Zhang, J., Chen, L., Chen, S.R., Chen, H., Zhang G & Pan, H.L. (2020). Histone methyltransferase G9a diminishes expression of cannabinoid CB1 receptors in primary sensory neurons in neuropathic pain. *Journal of Biological Chemistry, 295*(11), 3553-3562. https://doi.org/10.1074/jbc.ra119.011053

Muneer, A., Wang, L., Xie, L., Zhang, F., Wu, B., Mei, L., Lenarcic, E.M., Feng, E.H. et al. (2023). Non-canonical function of histone methyltransferase G9a in the translational regulation of chronic inflammation. *Cell Chemical Biology, 30*(12), 1525-1541.e7. https://doi.org/10.1016/j.chembiol.2023.09.012

Park, J., Lee, K., Kim, K., & Yi, S-J. (2022) The role of histone modifications: from neurodevelopment to neurodiseases. Signal Transduction and Targeted Therapy, 7, 217. https://doi.org/10.1038/s41392-022-01078-9

Pedregosa, F., Varoquaux, G., Gramfort. A, Michel, V., Thirion, B., Grisel, O. *et al.*, "Scikit-learn: Machine Learning in Python," JMLR, vol. 12, pp. 2825-2830, 2011, https://scikit-learn.org/stable/about.html

PubChem. (2009). Compound Summary. *National Institutes of Health*. https://pubchem.ncbi.nlm.nih.gov/compound/25150857#section=InChIKey Accessed 20 February 2025

PubChem. (2010). Aqueous Solubility from MLSMR Stock Solutions. *National Institutes of Health.* https://pubchem.ncbi.nlm.nih.gov/bioassay/1996 Accessed 20 February 2025

PubChem. (2011). qHTS Assay for Inhibitors of Histone Lysine Methyltransferase G9a. *National Institutes of Health.* https://pubchem.ncbi.nlm.nih.gov/bioassay/504332 Accessed 20 February 2025

PubChem (2024) Inhibition of G9a (unknown origin). *National Institutes of Health.* https://pubchem.ncbi.nlm.nih.gov/bioassay/1938431#section=Data-Table Accessed 20 February 2025

PubChem. (2025a). Explore Chemistry. *National Institutes of Health.* https://pubchem.ncbi.nlm.nih.gov/ Accessed 20 February 2025

PubChem. (2025b). About PubChem. *National Institutes of Health.* https://pubchem.ncbi.nlm.nih.gov/docs/about Accessed 20 February 2025

PubChem. (2025c). Compound Summary. *National Institutes of Health.* https://pubchem.ncbi.nlm.nih.gov/compound/171347753 Accessed 20 February 2025

Roopra, A., Qazi, R., Schoenike, B., Daley, T.J. & Morrison, J.F. (2004). Localized Domains of G9a-Mediated Histone Methylation Are Required for Silencing of Neuronal Genes. *Molecular Cell, 14*(6), 727-738. https://doi.org/10.1016/j.molcel.2004.05.026

Rothammer, N., Woo, M.S., Bauer, S., Binkle-Ladisch, L., Di Liberto, G., Egervari, K. et al. (2022) G9a dictates neuronal vulnerability to inflammatory stress via transcriptional control of ferroptosis. *Science Advances*, 8(31), eabm5500. https://doi.org/10.1126/sciadv.abm5500

Rukh, G., Akbar, S., Rehman, G., Alarfaj, F.K. and Zou, Q. (2024) StackedEnC-AOP: prediction of antioxidant proteins using transform evolutionary and sequential features based multi-scale vector with stacked ensemble learning. *BMC Bioinformatics,* 25, 256. https://doi.org/10.1186/s12859-024-05884-6

Shahid, Hayat, M., Alghamdi, W. Akabar, S., Raza, A., Kadir, R.A. and Sarker, M.R. (2025a) pACP-HybDeep: predicting anticancer peptides using binary tree growth based transformer and structural feature encoding with deep-hybrid learning. *Sci Rep*.,15, 565. https://doi.org/10.1038/s41598-024-84146-0

Shahid, Hayat, M., Raza, A., Akbar, S., Alghamdi, W., Iqbal, N., Zou, Q. (2025b) pACPs-DNN: Predicting anticancer peptides using novel peptide transformation into evolutionary and structure matrix-based images with self-attention deep learning model. *Comput Biol Chem*.,117, 108441. https://doi.org/10.1016/j.compbiolchem.2025.108441

Scikit-learn. (2025a). Home Page. *Scikit-learn.* https://scikit-learn.org/stable/index.html Accessed 20 February 2025

Scikit-learn (2025b) Feature importances with a forest of trees. *Scikit-learn.* https://scikit-learn.org/stable/auto_examples/ensemble/plot_forest_importances.html Accessed 20 February 2025

Scikit-learn (2025c) SelectKBest. *Scikit-learn.* https://scikit-learn.org/stable/modules/generated/sklearn.feature_selection.SelectKBest.html Accessed 20 February 2025

Singh G, Vadera M, Samavedham L, and Lim E.C-H. (2019). Multiclass Diagnosis of Neurodegenerative Diseases: A Neuroimaging Machine-Learning-Based Approach, *Industrial & Engineering Chemistry Research, 58* (26), 11498-11505. https://doi.org/10.1021/acs.iecr.8b06064

Ullah, M., Shahid Akbar, S., Ali Raza, A., Kashif Ahmad Khan, K.A., Quan Zou, Q. (2024) TargetCLP: clathrin proteins prediction combining transformed and evolutionary scale modelling-based multi-view features via weighted feature integration approach, *Briefings in Bioinformatics*, (26)1, bbaf026, https://doi.org/10.1093/bib/bbaf026

Wang, A.Y-T., Murdock R.J., Kauwe S.K., Oliynyk, A.O., Gurlo, A., Brgoch ,J., Kristin et al. (2020). Machine Learning for Materials Scientists: An Introductory Guide toward Best Practices. *Chemistry of Materials, 32* (12), 4954-4965. https://doi.org/10.1021/acs.chemmater.0c01907

Wang, M., He, A., Kang, Y., Wang, Z., He, Y., Lim, K., Zhang, C., & Lu, L. (2025) Novel genes involved in vascular dysfunction of the middle temporal gyrus in Alzheimer's disease: transcriptomics combined with machine learning analysis. *Neural Regeneration Research* 20(12), 3620-3634, https://doi.org/10.4103/nrr.nrr-d-23-02004

Wilson, C., Giono, L.E., Rozés-Salvador, V., Fiszbein, A., Kornblihtt, A.A., & Caceres, A. (2020) The Histone Methyltransferase G9a Controls Axon Growth by Targeting the RhoA Signalling Pathway. *Cell Reports*, 31(6), 107639. https://doi.org/10.1016/j.celrep.2020.107639

## Electronic Supplementary Material

### Figures

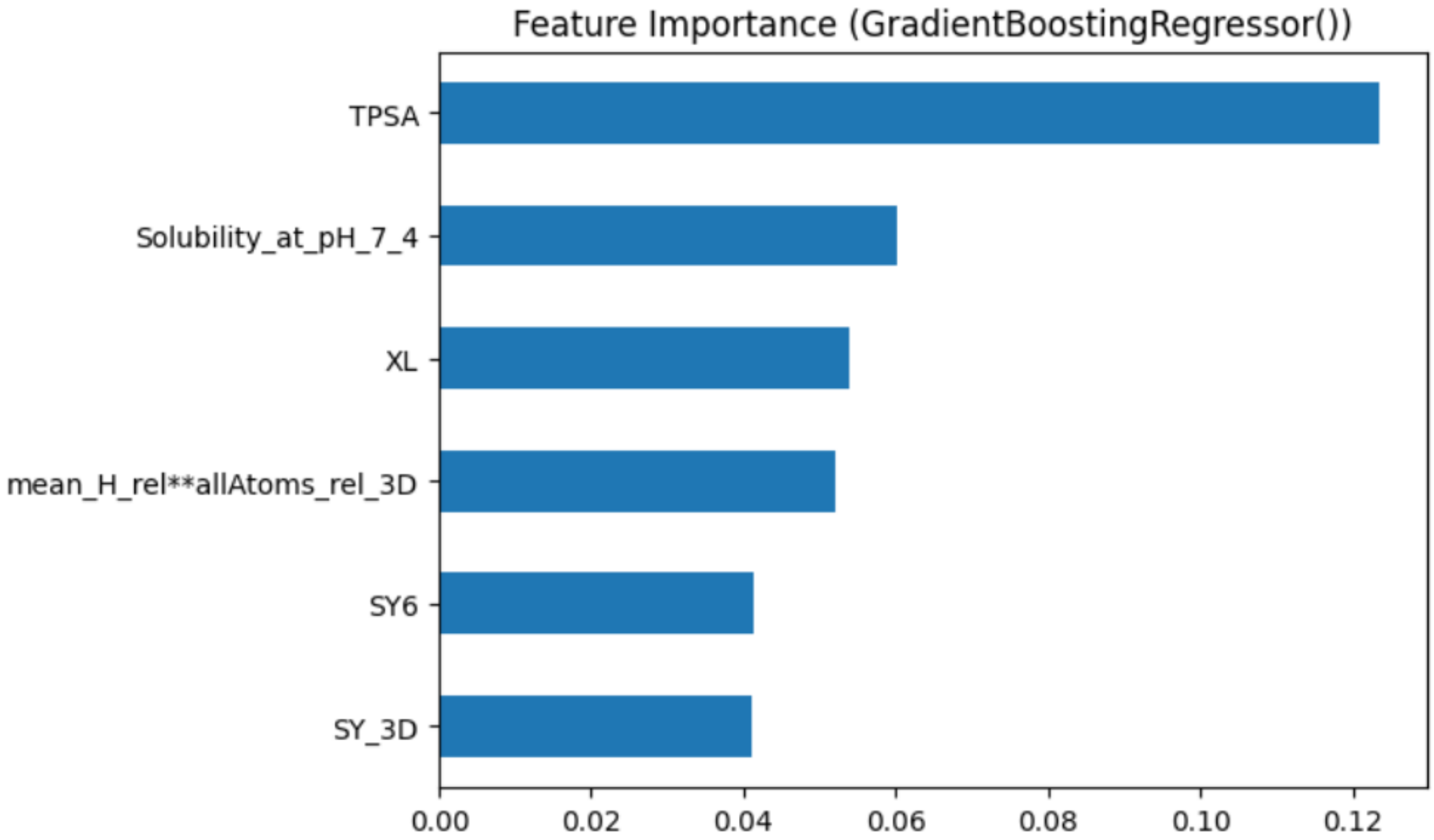

Figure ESM1. Feature importance of Gradient Boosting Regressor

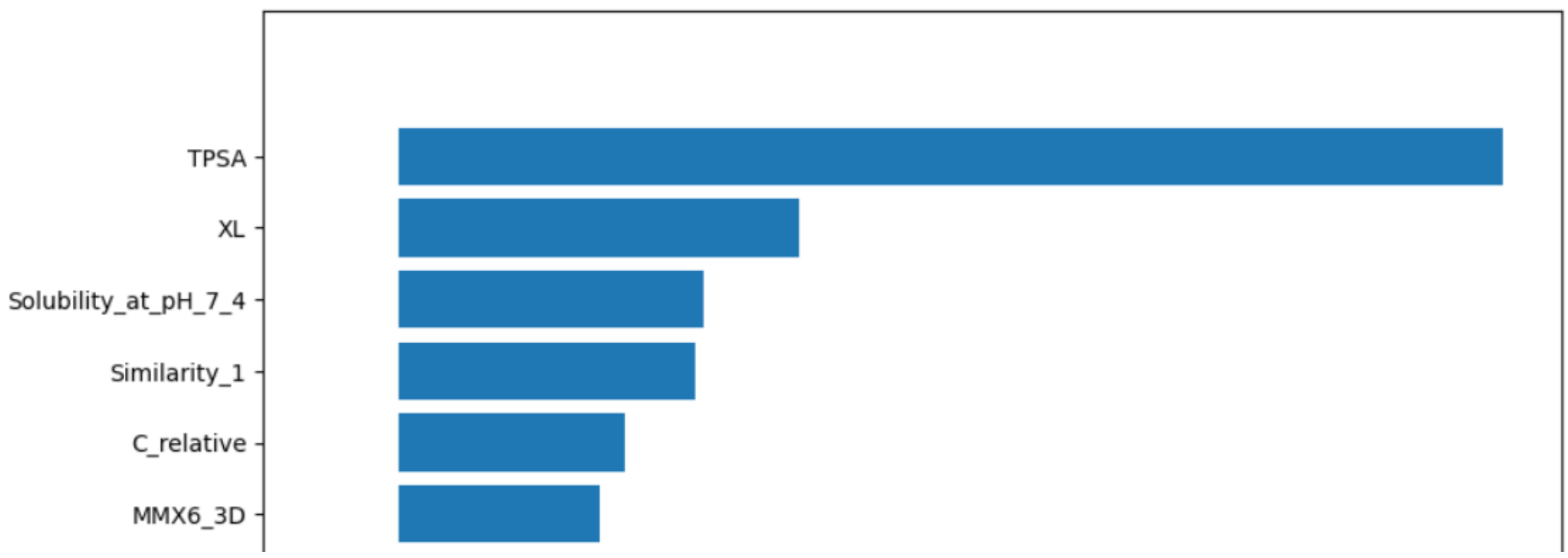

Figure ESM2. Permutation importance of the features train the Gradient Boosting Regressor

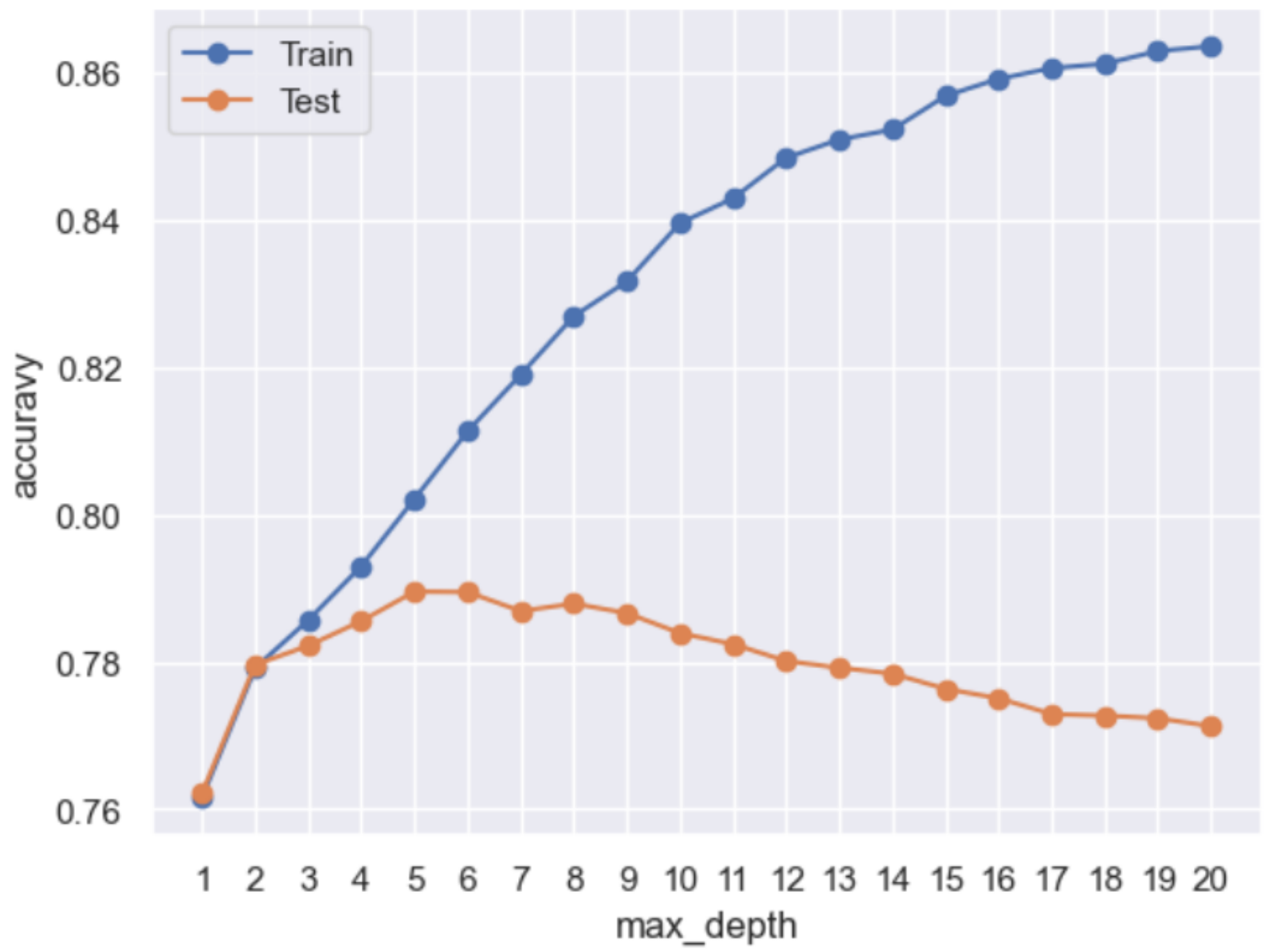

Figure ESM3. Scrutinized for overfitting of the CID_SID ML model with default hyperparameters

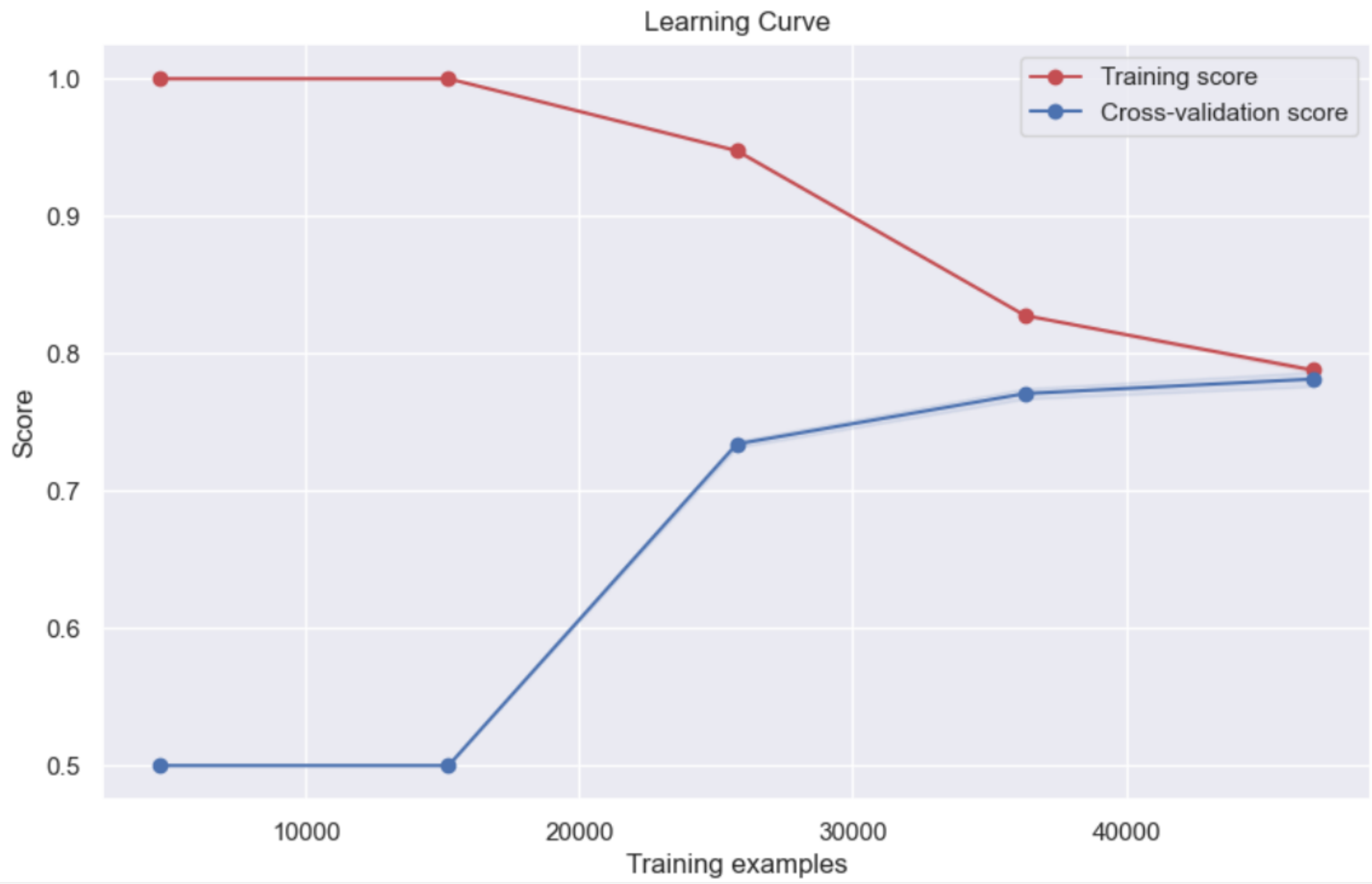

Figure ESM4. CID_SID ML model learning curves

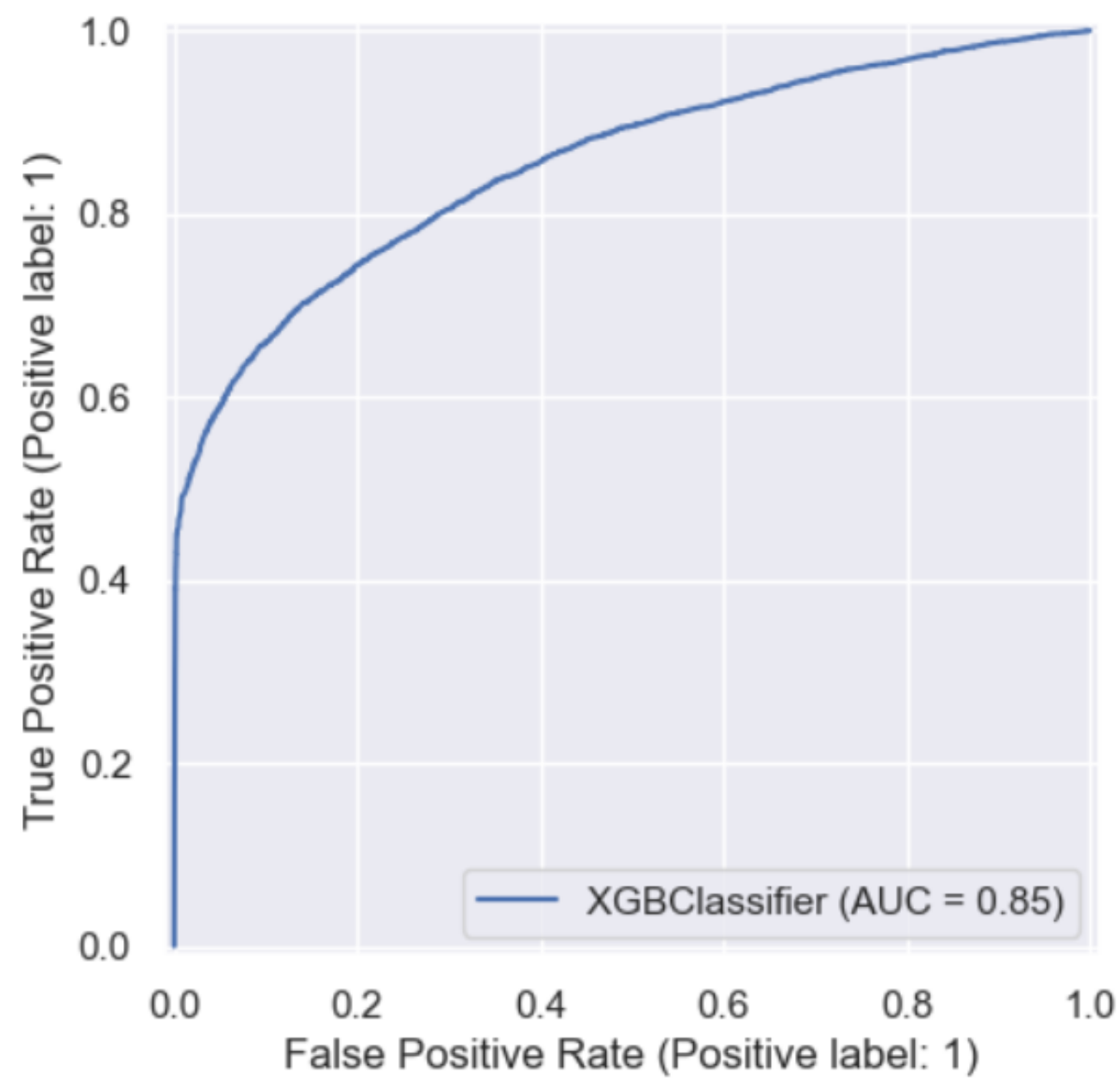

Figure ESM5. ROC of CID_SID ML model

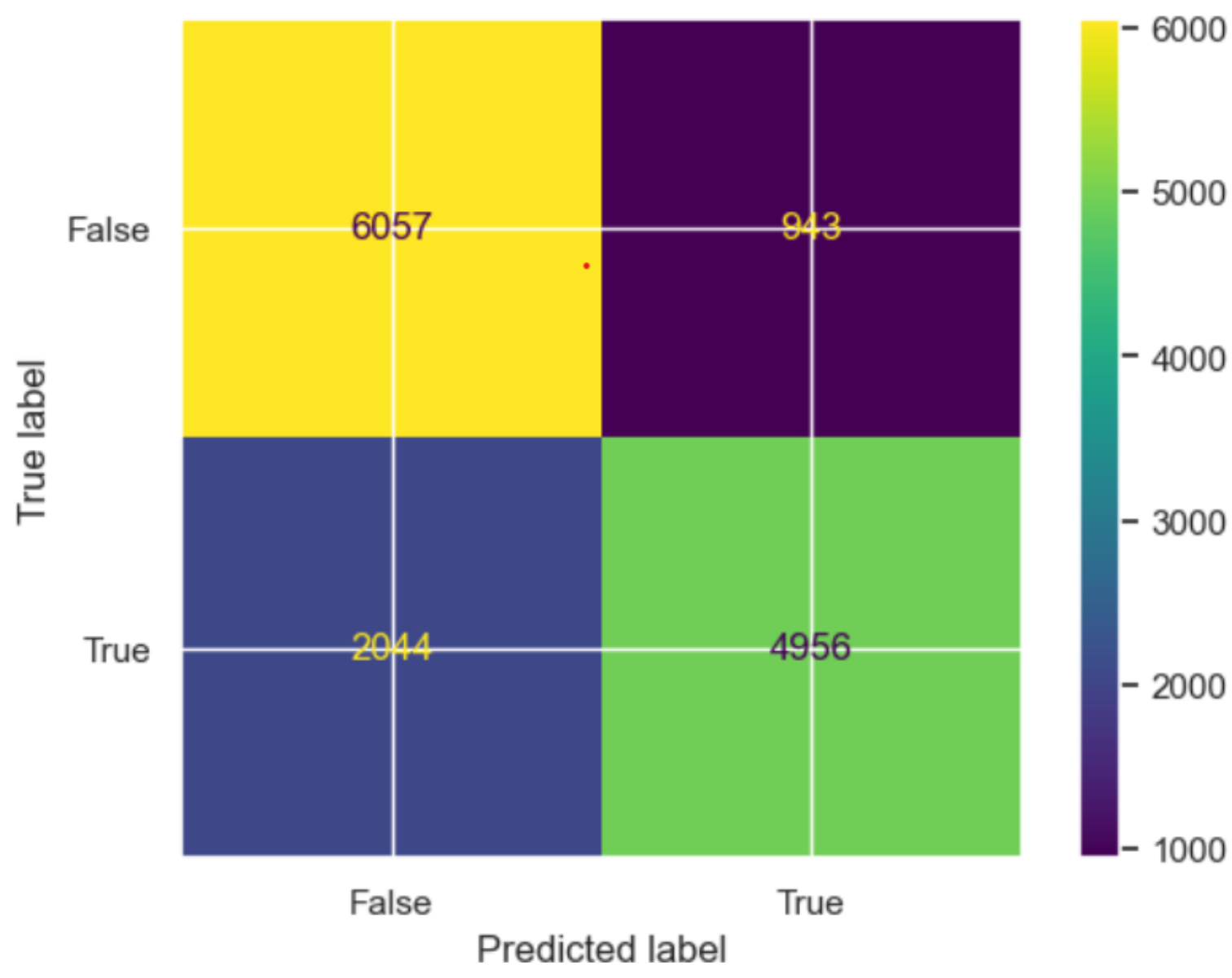

Figure ESM6. The CID_SID ML model confusion matrix

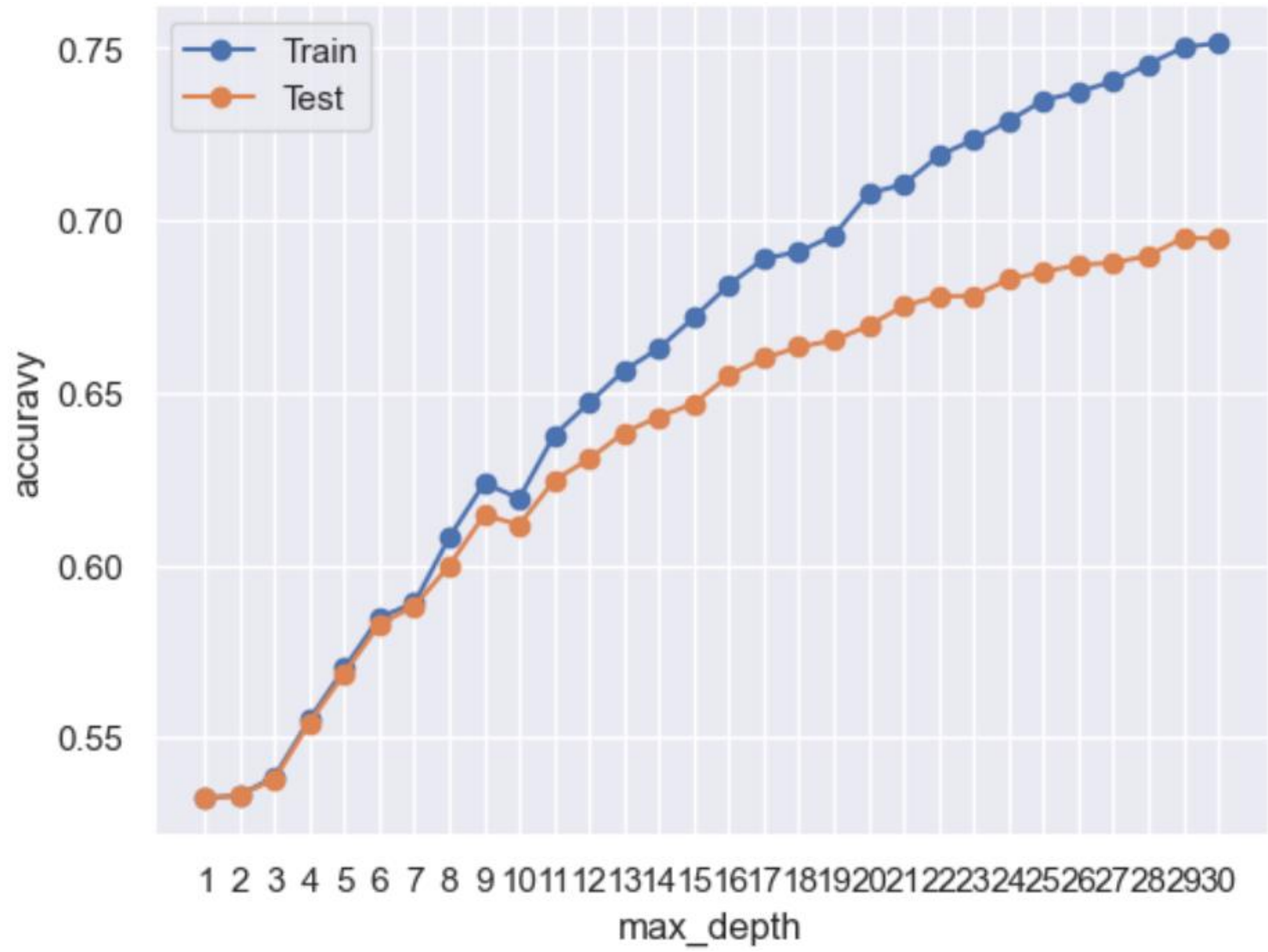

Figure ESM7. Scrutinizing for overfitting of IUPAC based Random Forest Classifier ML model

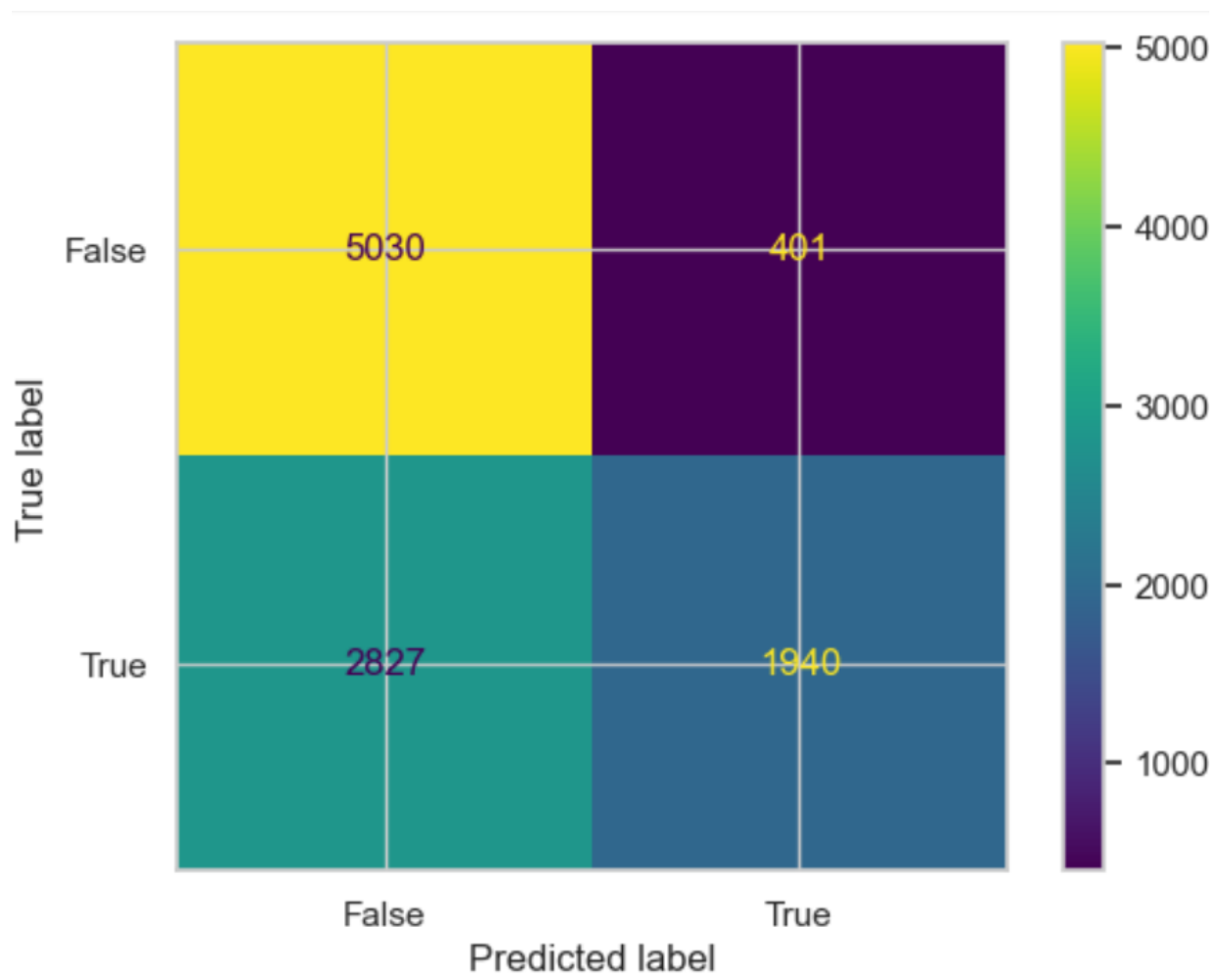

Figure ESM8. The IUPAC based Random Forest Classifier ML model confusion matrix

## Tables

Table ESM1. Evaluation of ML models that have been utilized Dataset basic with root mean squared error (MSE), mean absolute error (MAE) and R-squared (R2) metrics.

| Algorithm | RMSE_Train | RMSE_Test | MAE_Train | MAE_Test | R2_Train | R2_Test |
|---|---|---|---|---|---|---|
| RandomForest | 11.03 | 29.08 | 8.63 | 23.19 | 0.87 | 0.05 |
| GradientBoost | 28.89 | 29.29 | 22.84 | 23.37 | 0.09 | 0.04 |
| SVR | 29.88 | 29.50 | 23.53 | 23.55 | 0.03 | 0.03 |
| DecisionTree | 0.15 | 42.26 | 0.00 | 33.34 | 1.00 | -1.00 |

Table ESM2. Five-fold cross-validation of ML models that have been utilized Dataset basic

| Algorithm | CV_MAE | Sta Dev MAE | List of MAE |
|---|---|---|---|
| RandomForest | 23.31 | 0.26 | [23.17, 23.27, 23.71, 22.96, 23.46] |
| GradientBoost | 23.45 | 0.25 | [23.26, 23.37, 23.84, 23.15, 23.66] |
| SVR | 23.61 | 0.21 | [23.4, 23.63, 23.92, 23.35, 23.75] |
| DecisionTree | 33.45 | 0.28 | [33.14, 33.21, 33.83, 33.34, 33.76] |

Table ESM3 Evaluation of ML models that have been utilized Dataset (2) reduced with root mean squared error (MSE), mean absolute error(MAE) and R-squared (R2) metrics.

| Algorithm | RMSE_Train | RMSE_Test | MAE_Train | MAE_Test | R2_Train | R2_Test |
|---|---|---|---|---|---|---|
| RandomForest | 10.71 | 28.96 | 8.33 | 22.50 | 0.87 | 0.04 |
| GradientBoost | 24.55 | 29.18 | 19.37 | 22.71 | 0.29 | 0.02 |
| SVR | 28.54 | 29.78 | 22.28 | 23.13 | 0.05 | -0.02 |
| DecisionTree | 0.00 | 42.02 | 0.00 | 33.09 | 1.00 | -1.03 |

Table ESM4 Five-fold cross-validation of ML models that have been utilized Dataset (2) reduced

| Algorithm | CV_MAE | Sta Dev MAE | List of MAE |
|---|---|---|---|
| RandomForest | 22.43 | 0.12 | [22.54, 22.23, 22.53, 22.35, 22.5] |
| GradientBoost | 22.47 | 0.30 | [22.74, 22.0, 22.75, 22.23, 22.63] |
| SVR | 22.77 | 0.33 | [23.34, 22.37, 22.93, 22.64, 22.57] |
| DecisionTree | 31.78 | 1.23 | [32.18, 33.15, 29.92, 32.84, 30.82] |

Table ESM5. Evaluation of ML models that have been utilized Dataset (3) extended with root mean squared error (MSE), mean absolute error(MAE) and R-squared (R2) metrics.

| Algorithm | RMSE_Train | RMSE_Test | MAE_Train | MAE_Test | R2_Train | R2_Test |
|---|---|---|---|---|---|---|
| RandomForest | 10.69 | 28.99 | 8.37 | 22.47 | 0.87 | 0.04 |
| GradientBoost | 24.54 | 29.29 | 19.46 | 22.80 | 0.30 | 0.02 |
| SVR | 28.43 | 29.89 | 22.18 | 23.32 | 0.05 | -0.03 |
| DecisionTree | 0.00 | 40.39 | 0.00 | 32.37 | 1.00 | -0.87 |

Table ESM6. Five-fold cross-validation of ML models that have been utilized Dataset (3) extended with hew features

| Algorithm | CV_MAE | Sta Dev MAE | List of MAE |
|---|---|---|---|
| RandomForest | 22.55 | 0.24 | [22.85, 22.15, 22.74, 22.49, 22.5] |
| GradientBoost | 22.61 | 0.34 | [22.92, 22.05, 22.96, 22.4, 22.75] |
| SVR | 22.74 | 0.34 | [23.33, 22.46, 22.9, 22.56, 22.46] |
| DecisionTree | 32.14 | 0.61 | [33.03, 31.64, 31.31, 32.46, 32.27] |

Table ESM7 Evaluation of ML models that have been utilized Dataset (4) no skewness with root mean squared error (MSE), mean absolute error (MAE) and R-squared (R2) metrics.

| Algorithm | RMSE_Train | RMSE_Test | MAE_Train | MAE_Test | R2_Train | R2_Test |
|---|---|---|---|---|---|---|
| GradientBoost | 25.51 | 27.03 | 20.17 | 21.29 | 0.25 | 0.08 |
| RandomForest | 10.95 | 27.38 | 8.55 | 21.42 | 0.86 | 0.06 |
| SVR | 29.14 | 27.75 | 22.77 | 21.97 | 0.03 | 0.03 |
| DecisionTree | 0.00 | 38.98 | 0.00 | 30.45 | 1.00 | -0.91 |

Table ESM8 Five-fold cross-validation of ML models that have been utilized Dataset (4) without skewness

| Algorithm | CV_MAE | Sta Dev MAE | List of MAE |
|---|---|---|---|
| RandomForest | 22.94 | 0.70 | [23.46, 21.7, 22.91, 23.74, 22.87] |
| GradientBoost | 23.06 | 0.85 | [23.67, 21.52, 23.35, 23.93, 22.85] |
| SVR | 23.14 | 0.68 | [23.5, 21.95, 23.2, 24.02, 23.02] |
| DecisionTree | 33.07 | 0.89 | [34.68, 32.69, 32.58, 33.32, 32.09] |

Table ESM9. Evaluation of ML models that have been utilized Dataset (5) without skewness after log transformation with root mean squared error (MSE), mean absolute error(MAE) and R-squared (R2) metrics.

| Algorithm | RMSE_Train | RMSE_Test | MAE_Train | MAE_Test | R2_Train | R2_Test |
|---|---|---|---|---|---|---|
| RandomForest | 11.07 | 27.75 | 8.62 | 21.92 | 0.86 | 0.03 |
| GradientBoost | 26.11 | 27.78 | 20.61 | 21.94 | 0.22 | 0.03 |
| SVR | 29.38 | 27.98 | 23.02 | 22.05 | 0.01 | 0.01 |
| DecisionTree | 0.00 | 39.81 | 0.00 | 31.80 | 1.00 | -0.99 |

Table ESM10. Five-fold cross-validation of ML models that have been utilized Dataset (5) without skewness after log transformation

| Algorithm | CV_MAE | Sta Dev MAE | List of MAE |
|---|---|---|---|
| SVR | 23.33 | 0.71 | [23.89, 22.33, 23.28, 24.33, 22.84] |
| RandomForest | 23.34 | 0.69 | [23.96, 22.14, 23.24, 24.08, 23.27] |
| GradientBoost | 23.64 | 0.64 | [24.32, 22.82, 23.37, 24.47, 23.23] |
| DecisionTree | 33.28 | 0.51 | [33.8, 32.76, 33.19, 32.72, 33.93] |

Table ESM11 Five-fold cross-validation of CID_SID ML models

| Algorithm | Mean CV Score | Standard Deviation | List of CV Scores |
|---|---|---|---|
| XGBoost | 0.7998 | 0.0013 | [0.7981, 0.7986, 0.8006, 0.8, 0.8015] |
| GradientBoost | 0.7912 | 0.0036 | [0.7866, 0.7901, 0.7976, 0.7906, 0.7913] |
| RandomForest | 0.7908 | 0.0015 | [0.7885, 0.7897, 0.7918, 0.7921, 0.7922] |
| K-nearest | 0.7777 | 0.0033 | [0.7725, 0.7754, 0.7804, 0.7813, 0.7789] |
| Decision | 0.7547 | 0.0033 | [0.7525, 0.7544, 0.7559, 0.7504, 0.7603] |
| SVM | 0.7429 | 0.0036 | [0.7385, 0.7393, 0.748, 0.7454, 0.7434] |